\documentclass[longauth]{aa}
\usepackage{graphicx}
\usepackage{longtable,lscape}
\usepackage{epsfig}
\usepackage{amsmath}

\begin{document}
\title{Spectrum and variability of the Galactic Center VHE $\gamma$-ray source HESS J1745$-$290}

\small{
\author{F. Aharonian\inst{1,13}
 \and A.G.~Akhperjanian \inst{2}
 \and G.~Anton \inst{16}
 \and U.~Barres de Almeida \inst{8} \thanks{supported by CAPES Foundation, Ministry of Education of Brazil}
 \and A.R.~Bazer-Bachi \inst{3}
 \and Y.~Becherini \inst{12}
 \and B.~Behera \inst{14}
 \and K.~Bernl\"ohr \inst{1,5}
 \and C.~Boisson \inst{6}
 \and A.~Bochow \inst{1}
% ...
}
}

\offprints{Matthieu Vivier \email{matthieu.vivier@cea.fr}, Christopher van Eldik \email{Christopher.van.Eldik@mpi-hd.mpg.de}}

\institute{
Max-Planck-Institut f\"ur Kernphysik, P.O. Box 103980, D 69029
Heidelberg, Germany
\and
 Yerevan Physics Institute, 2 Alikhanian Brothers St., 375036 Yerevan,
Armenia
\and
Centre d'Etude Spatiale des Rayonnements, CNRS/UPS, 9 av. du Colonel Roche, BP
4346, F-31029 Toulouse Cedex 4, France
\and
Universit\"at Hamburg, Institut f\"ur Experimentalphysik, Luruper Chaussee
149, D 22761 Hamburg, Germany
\and
Institut f\"ur Physik, Humboldt-Universit\"at zu Berlin, Newtonstr. 15,
D 12489 Berlin, Germany
\and
LUTH, Observatoire de Paris, CNRS, Universit\'e Paris Diderot, 5 Place Jules Janssen, 92190 Meudon, 
France
Obserwatorium Astronomiczne, Uniwersytet Ja
\and
IRFU/DSM/CEA, CE Saclay, F-91191
Gif-sur-Yvette, Cedex, France
\and
University of Durham, Department of Physics, South Road, Durham DH1 3LE,
U.K.
\and
Unit for Space Physics, North-West University, Potchefstroom 2520,
    South Africa
\and
Laboratoire Leprince-Ringuet, Ecole Polytechnique, CNRS/IN2P3,
 F-91128 Palaiseau, France
\and 
Laboratoire d'Annecy-le-Vieux de Physique des Particules, CNRS/IN2P3,
9 Chemin de Bellevue - BP 110 F-74941 Annecy-le-Vieux Cedex, France
\and
Astroparticule et Cosmologie (APC), CNRS, Universite Paris 7 Denis Diderot,
10, rue Alice Domon et Leonie Duquet, F-75205 Paris Cedex 13, France
\thanks{UMR 7164 (CNRS, Universit\'e Paris VII, CEA, Observatoire de Paris)}
\and
Dublin Institute for Advanced Studies, 5 Merrion Square, Dublin 2,
Ireland
\and
Landessternwarte, Universit\"at Heidelberg, K\"onigstuhl, D 69117 Heidelberg, Germany
\and
Laboratoire de Physique Th\'eorique et Astroparticules, CNRS/IN2P3,
Universit\'e Montpellier II, CC 70, Place Eug\`ene Bataillon, F-34095
Montpellier Cedex 5, France
\and
Universit\"at Erlangen-N\"urnberg, Physikalisches Institut, Erwin-Rommel-Str. 1,
D 91058 Erlangen, Germany
% ...
}

\authorrunning{F. Aharonian et al.}
\titlerunning{Spectrum and variability of HESS J1745-290}
\date{June 30, 2009}

\abstract{}
{A detailed study of the spectrum and variability of the source HESS J1745$-$290 in the Galactic Center (GC) region using new data from the H.E.S.S. array of Cherenkov telescopes is presented. Flaring activity and quasi periodic oscillations (QPO) of HESS J1745$-$290 are investigated.}
{The image analysis is performed with a combination of a semi-analytical shower model and the statistical moment-based Hillas technique. The spectrum and lightcurves of HESS J1745$-$290 are derived with a likelihood method based on a spectral shape hypothesis. Rayleigh tests and Fourier analysis of the H.E.S.S. GC signal are used to study the periodicity of the source.}
{With a three-fold increase in statistics compared to previous work, a deviation from a simple power law spectrum is detected for the first time. The measured energy spectrum over the three years 2004, 2005 and 2006 of data taking is compatible with both a power law spectrum with an exponential cut-off and a broken power law spectrum. The curvature of the energy spectrum is likely to be intrinsic to the photon source, as opposed to effects of interstellar absorption. The power law spectrum with an exponential cut-off is characterized by a photon index of 2.10 $\pm$ 0.04$_{\mathrm{stat}}$ $\pm$ 0.10$_{\mathrm{syst}}$ and a cut-off energy at 15.7 $\pm$ 3.4$_{\mathrm{stat}}$ $\pm$ 2.5$_{\mathrm{syst}}$ TeV. The broken power law spectrum exhibits spectral indices of 2.02 $\pm$ 0.08$_{\mathrm{stat}}$ $\pm$ 0.10$_{\mathrm{syst}}$ and 2.63 $\pm$ 0.14$_{\mathrm{stat}}$ $\pm$ 0.10$_{\mathrm{syst}}$ with a break energy at 2.57 $\pm$ 0.19$_{\mathrm{stat}}$ $\pm$ 0.44$_{\mathrm{syst}}$ TeV. No significant flux variation is found. Increases in the $\gamma$-ray flux of HESS J1745$-$290 by at least a factor of two would be required for a 3$\sigma$ detection of a flare with time scales of an hour. Investigation of possible QPO activity at periods claimed to be detected in X-rays does not show any periodicities in the H.E.S.S. signal.}
{}


\maketitle


\section{Introduction}
The discovery of very high energy (VHE) $\gamma$-rays from the Galactic Center (GC) has been reported by CANGAROO (Tsuchiya et al \cite{Tsuchiya04}), VERITAS (Kosack et al. \cite{Kosack}), H.E.S.S. (HESS J1745$-$290, Aharonian et al. \cite{Aharonian2}) and MAGIC (Albert et al. \cite{Albert}). Possible associations of the source with the Sgr A East supernova remnant (Crocker et al. \cite{Crocker}) and more recently with the newly detected plerion G359.95$-$0.04 (Wang et al. \cite{Wang}) have been widely discussed in the literature. However, with the reduced systematic pointing error obtained using H.E.S.S. data up to 2006 (van Eldik et al. \cite{VanEldik}), Sgr A East is now ruled out as being associated with the VHE emission of HESS J1745$-$290. The interpretation of the GC TeV signal as annihilation products of dark matter (DM) particles has been discussed in Aharonian et al. (\cite{PRL}). It is unlikely that the bulk of the signal comes from DM annihilations. One possibility is that the supermassive black hole Sgr A* located at the center of the Milky Way is responsible for the VHE emission of the detected HESS J1745$-$290 source. In this case, a cut-off in the high energy part of the spectrum might be expected (Ballantyne et al. \cite{Ballantyne}, Aharonian and Neronov \cite{Neronov1}). Time variability as seen in X-rays and IR might also appear, along with flares, and QPO frequencies such as those found in Aschenbach et al. (\cite{Aschenbach}). QPO periods of $\approx$ 100 s, 219 s, 700 s, 1150 s and 2250 s have been observed simultaneously in two different datasets collected with the Chandra (Baganoff et al. \cite{Baganoff}) and the XMM-Newton (Porquet et al. \cite{Porquet}) observatories in 2000 and 2002, respectively. The power density spectra of the 2003 infrared flare also shows three distinct peaks at $\approx$ 214 s, 733 s and 1026 s (Genzel et al. \cite{Genzel}), fully consistent with three of the five X-ray detected periods. Recently however, the validity of the detection of these periods has been disputed by infrared observations on the Keck II telescope (see Meyer et al. (\cite{Meyer}) for references).\\
The search for a curvature in the TeV energy spectrum and time variability in the TeV signal is strongly motivated by GC wideband emission models and multi-wavelength data, respectively. In this paper, updated results on the energy spectrum and variability of HESS J1745$-$290 are presented, based on a dataset collected in 2004, 2005 and 2006. Preliminary results on the source position and morphology were published in van Eldik et al. (\cite{VanEldik}). 

\section{H.E.S.S. observations and analysis}
\subsection{The H.E.S.S. instrument}
The H.E.S.S. (High Energy Stereoscopic System) instrument (Hofmann et al. \cite{Hofmann}) consists of four Imaging Atmospheric Cherenkov Telescopes (IACTs) located in the Khomas Highland of Namibia at an altitude of 1800 m above sea level. H.E.S.S. is dedicated to very high energy $\gamma$-ray astronomy (defined as E $\geq$ 100 GeV), beyond the energy range accessible to satellite-based detectors. The IACTs have a large mirror area of 107 m$^{2}$, reflecting the Cherenkov light emitted by $\gamma$-induced air showers onto a camera of 960 photomultiplier tubes (PMTs). Each PMT covers a field of view of 0.16\degr. The total field of view of the camera is 5\degr in diameter. The telescopes are positioned at the corners of a square, of side 120 m, which allows for an accurate reconstruction of the direction and energy of the $\gamma$-rays using the stereoscopic technique. The energy threshold of H.E.S.S. is approximately 100 GeV at zenith. More details on the H.E.S.S. instrument can be found in Aharonian et al. (\cite{Crab}).

\subsection{Dataset and analysis}
\begin{figure}[htbp]
\begin{center}
\resizebox{\hsize}{!}{\includegraphics {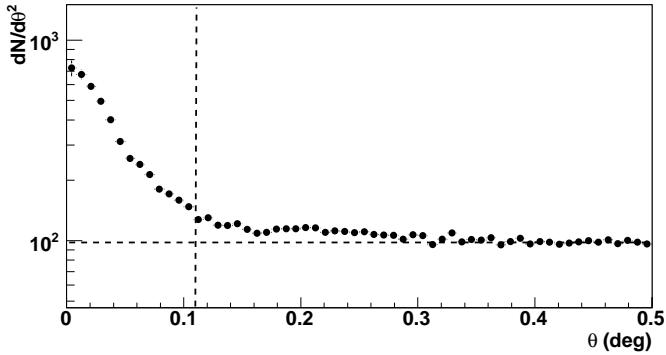}}
\end{center}
\vspace{-0.3cm}\caption{Distribution of the number of $\gamma$-rays derived with the combined Hillas/Model analysis as a function of $\theta$, where $\theta$ is the angle between the $\gamma$-ray direction and the GC position. The dashed vertical line delimits the ON-source region. The horizontal line shows the expected number of cosmic ray background events.}\label{fig6}
\end{figure}

The previously published H.E.S.S. observations on the GC (Aharonian et al. \cite{PRL}) was on a 48.7 hours (live time) data sample collected in 2004. In this paper new results on the spectral analysis of the GC are derived, using data from subsequent observation campaigns carried out in 2005 and 2006, with zenith angles ranging up to 70\degr. The runs taken at high zenith angles are sensitive to higher $\gamma$-ray energies and allow us to probe the high energy part of the HESS J1745$-$290 spectrum. The datasets are described in Table \ref{table1}.
\begin{table}[htp]
\renewcommand\footnoterule{} 
\begin{minipage}[t]{90mm}
\caption{Details of the observation of the GC region with H.E.S.S. for the 2004, 2005 and 2006 observation campaigns.}
\label{table1}
\begin{center}
\begin{tabular}{|c|c|c|c|c|c|}
\hline
Year & $\theta_z$\footnote{$\theta_z$ denotes the zenith angle} range & $\bar{\theta_z}^{a}$ & T$_{obs}$\footnote{Live time of the data sample} & $\rm N_{\gamma}$\footnote{Number of detected $\gamma$-rays after background substraction} & $\sigma$\footnote{Significance of the $\gamma$-ray excess N$_{\gamma}$ calculated according to the prescription of Li and Ma (\cite{LiMa})}\\
& (\degr) & (\degr) & (hrs) & & \\
\hline 
2004 & 0-60 & 21.3 & 28.5 & 1516 & 35.6 \\
2005 & 0-70 & 26.6 & 51.7 & 2062 & 41.2 \\
2006 & 0-50 & 19.1 & 12.7 & 607 & 23.0 \\
\hline
\hline
All  & 0-70 & 23.0 & 92.9 & 4185 & 60.7 \\
\hline
\end{tabular}
\end{center}
\end{minipage}
\end{table}
Most of the data were taken in ``wobble mode'' where the telescope pointing is typically shifted by $\pm$ 0.7\degr from the nominal target position. The dataset used for the analysis was selected using the standard quality criteria, excluding runs taken under bad and variable weather conditions, which would lead to large systematic errors. An additional quality selection was applied using the cosmic ray rate during data collection. Runs with rates deviating by more than 3$\sigma$ from the cosmic ray rate averaged over the three years of observation were removed in the subsequent analysis. After the run selection procedure, the dataset amounts to 93 hours of live time.\\
Two independent techniques are commonly used to select $\gamma$-ray events and derive energy spectra and lightcurves. The first technique computes the ``Hillas geometrical moments'' of the shower image (Aharonian et al. \cite{Hillas}) and the second one is based on a semi-analytical model of showers, which predicts the expected intensity in each pixel of the camera (de Naurois et al. \cite{deNaurois}). The shower direction, the impact point and the primary particle energy are then derived with a likelihood fit of the model to match the data. Both analysis techniques provide an energy resolution of 15-20$\%$ with an angular resolution better than 0.1\degr above the analysis energy threshold. Results described in this paper are obtained with a combination of these two analysis techniques, the so-called combined Hillas/Model analysis (de Naurois et al. \cite{deNaurois}), which yields an improved background rejection, and were cross-checked against either technique alone. The background is calculated at each position in the field of view using the acceptance corrected background rate from an annulus around that position (the OFF-source region). More details on this so-called ring background method are given in Berge et al. (\cite{Berge}). The OFF-source regions do not overlap with known TeV sources in the GC field of view (Aharonian et al. \cite{GalacticScanPaper}) and areas of diffuse emission (Aharonian et al. \cite{DE}).\\
The analysis of the whole 2004-2006 dataset shows an excess above the background of 4185 $\gamma$-events in a 0.11\degr radius region centered on the GC (Table \ref{table1}). The total significance, calculated according to the method of Li and Ma (\cite{LiMa}), is 60.7$\sigma$. Diffuse $\gamma$-ray emission along the galactic plane is visible as shown by the distribution in $\theta$, $\theta$ being the angle of a reconstructed $\gamma$-ray relative to Sgr A* (Fig. \ref{fig6}, excess above the background level outside the ON source region). A discussion in Aharonian et al. (\cite{DE}) gives a possible interpretation of the emission as cosmic ray interactions in the central molecular zone (CMZ). A point-source + linear background fit to the $\theta$ distribution has been performed to estimate the diffuse emission contribution to the signal. Extrapolating the linear component inside the ON-source region gives a diffuse emission contribution of 13$\%$ $\pm$ 1$\%$ within 0.11\degr. Diffuse $\gamma$-ray emission located in the ON-source region was not subtracted in the following analysis, since its spectral feature (a simple power law with the same spectral index as the central source) would not influence the result (Aharonian et al. \cite{DE}). Any deviation from a power law spectrum or variability in the lightcurve would be solely due to the point source. 
\section{Energy spectrum of HESS J1745$-$290}
\subsection{Spectral reconstruction}
The energy spectrum of HESS J1745$-$290 has been derived following the forward folding method. The forward folding method is based on a spectral shape hypothesis for the source spectrum (Piron et al. \cite{Piron1}, Djannati-Atai et al. \cite{Djannati}). Spectral points are then calculated using the adjusted spectral shape and the 1$\sigma$ error bars are computed using the error matrix of the fit procedure. Relative systematic errors on the reconstructed spectral indices coming from the presence of broken pixels in the camera are less than 5$\%$, whereas those coming from variations of the atmospheric conditions are negligible. Thus, systematic errors on the spectral indices derived in the following analysis are taken to be 5$\%$. Systematic errors on the integrated fluxes mainly come from the variations of the atmospheric conditions and the absolute calibration of the response of the telescopes. They amount to 10 to 20$\%$ (Aharonian et al. \cite{Crab}). The systematic errors on the integrated fluxes above 1 TeV are taken to be 20$\%$. A Monte-Carlo study of the reconstruction of the cut-off energy revealed a systematic bias linearly increasing with the cut-off energy:
\begin{equation}
\mathrm{E_{cut}} = (0.92 \pm 0.01) \times \mathrm{E_{cut,true}} + (0.25 \pm 0.05)\,\mathrm{TeV}.
\end{equation}
The systematic errors on the reconstruction of the cut-off energy amount to 17$\%$.
\subsection{Results}
\begin{figure*}[]
\begin{center}
\noindent
\mbox{\hspace{-0.3cm}\includegraphics [height=0.58\textwidth]{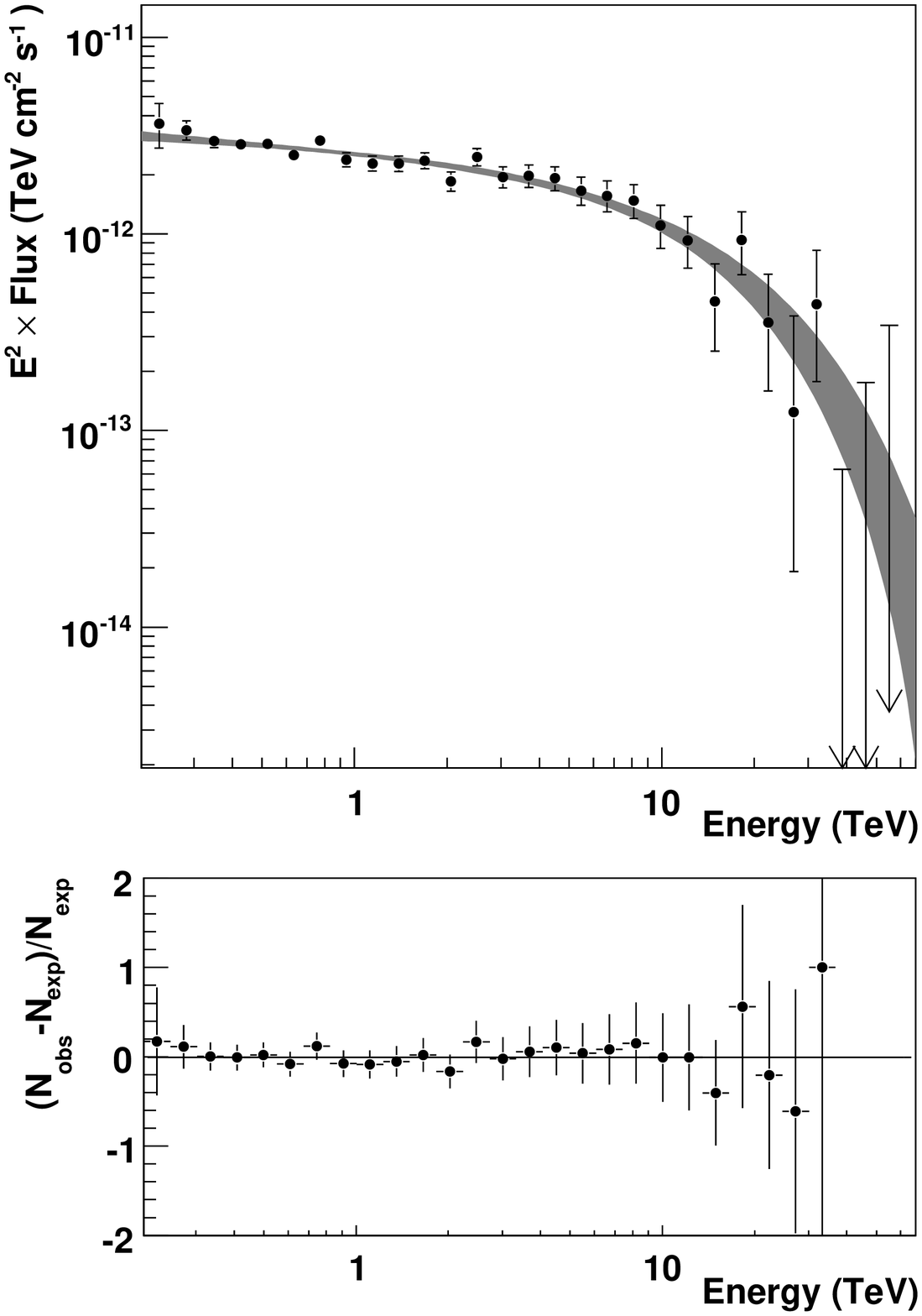} 
\hspace{-0.3cm}\includegraphics [height=0.58\textwidth]{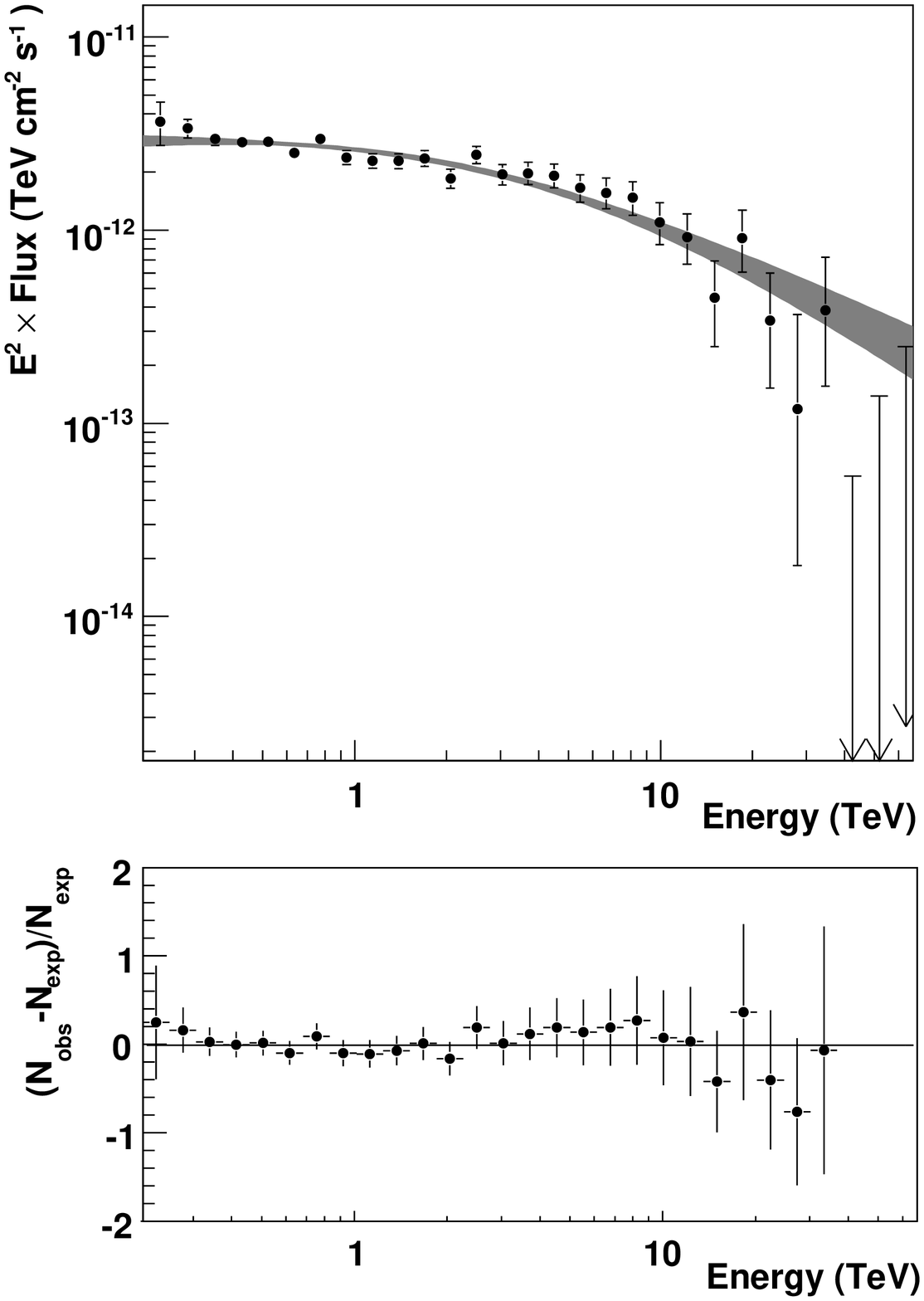}}
\end{center}
\vspace{-0.4cm}\caption{HESS J1745$-$290 spectra derived with the combined Hillas/Model analysis for the whole H.E.S.S. GC dataset covering the three years 2004, 2005 and 2006. The shaded areas are the 1$\sigma$ confidence intervals for the power law with an exponential cut-off fit (left) and the smoothed broken power law fit (right). The last points represent 95$\%$ confidence level upper limits on the flux. The fit residuals corresponding to the respective fits are shown on the lower panels.}\label{fig7}
\end{figure*}

The data taken in 2004, 2005 and 2006 were compared to the folding of three distributions with the detector response: a power law (Eq. \ref{Eq1}), a power law with a high energy exponential cut-off (Eq. \ref{Eq2}) and a smoothed broken power law (Eq. \ref{Eq3}):
\begin{align}
\frac{dN}{dE}&=\Phi_{0}\times\Big(\frac{E}{1 \rm{TeV}}\Big)^{-\Gamma}\label{Eq1}\\
\frac{dN}{dE}&=\Phi_{0}\times\Big(\frac{E}{1 \rm{TeV}}\Big)^{-\Gamma}\times e^{-(\frac{E}{E_{\rm{cut}}})^{\beta}}\label{Eq2}\\
\frac{dN}{dE}&=\Phi_{0}\times\Big(\frac{E}{1 \rm{TeV}}\Big)^{-\Gamma_1}\times\frac{1}{\Big(1+\Big(\frac{E}{E_{\rm{break}}}\Big)^{(\Gamma_2-\Gamma_1)}\Big)}\label{Eq3}
\end{align}
where $\Phi_{0}$ is the flux normalisation in TeV$^{-1}$ cm$^{-2}$ s$^{-1}$, $\Gamma_i$ the spectral indices. E$_{\rm{cut}}$ is the cut-off energy in Eq. \ref{Eq2}, and E$_{\rm{break}}$ is the break energy in Eq. \ref{Eq3}. In Eq. \ref{Eq2}, $\beta$ is the strength of the cut-off. Except at the end of paragraph 3.2.1., $\beta$ is taken equal to one.\\
The measured spectrum for the whole three-year dataset ranges from 160 GeV, the energy threshold of the analysis, to 70 TeV (Fig. \ref{fig7}). For the first time, with additional statistics, a deviation from a pure power law starts to be visible. The spectrum is well described by either Eq. \ref{Eq2} (equivalent $\chi^{2}$ of 23/26 d.o.f.) or Eq. \ref{Eq3} (equivalent $\chi^{2}$ of 20/19 d.o.f.). Fig. \ref{fig7} shows the HESS J1745$-$290 spectra with fits to Eq. \ref{Eq2} and Eq. \ref{Eq3}. The power law with an exponential cut-off fit yields $\Phi_0$ = (2.55 $\pm$ 0.06$_{\mathrm{stat}}$ $\pm$ 0.40$_{\mathrm{syst}}$) $\times$ 10$^{-12}$ TeV$^{-1}$ cm$^{-2}$ s$^{-1}$, $\Gamma$ = 2.10 $\pm$ 0.04$_{\mathrm{stat}}$ $\pm$ 0.10$_{\mathrm{syst}}$, a cut-off energy E$_{\rm{cut}}$ = (15.7 $\pm$ 3.4$_{\mathrm{stat}}$ $\pm$ 2.5$_{\mathrm{syst}}$) TeV and an integrated flux above 1 TeV of (1.99 $\pm$ 0.09$_{\mathrm{stat}}$ $\pm$ 0.40$_{\mathrm{syst}}$) $\times$ 10$^{-12} $cm$^{-2}$ s$^{-1}$. The broken power law fit yields $\Phi_0$ = (2.57 $\pm$ 0.07$_{\mathrm{stat}}$ $\pm$ 0.40$_{\mathrm{syst}}$) $\times$ 10$^{-12}$ TeV$^{-1}$ cm$^{-2}$ s$^{-1}$, $\Gamma_1$ = 2.02 $\pm$ 0.08$_{\mathrm{stat}}$ $\pm$ 0.10$_{\mathrm{syst}}$, $\Gamma_2$ = 2.63 $\pm$ 0.14$_{\mathrm{stat}}$ $\pm$ 0.10$_{\mathrm{syst}}$, a break energy E$_{\rm{break}}$ = (2.57 $\pm$ 0.19$_{\mathrm{stat}}$ $\pm$ 0.44$_{\mathrm{syst}}$) TeV and an integrated flux above 1 TeV of (1.98 $\pm$ 0.38$_{\mathrm{stat}}$ $\pm$ 0.40$_{\mathrm{syst}}$) $\times$ 10$^{-12} $cm$^{-2}$ s$^{-1}$. The values of the cut-off energy E$_{\rm{cut}}$ and break energy  E$_{\rm{break}}$ are those corrected for the systematic bias mentioned in the previous section. By comparison, a power law spectrum gives a worse equivalent $\chi^{2}$ of 64/27 d.o.f. and yields a flux normalisation of (2.40 $\pm$ 0.05$_{\mathrm{stat}}$ $\pm$ 0.40$_{\mathrm{syst}}$) $\times$ 10$^{-12}$ TeV$^{-1}$ cm$^{-2}$ s$^{-1}$, a spectral index of 2.29 $\pm$ 0.02$_{\mathrm{stat}}$ $\pm$ 0.10$_{\mathrm{syst}}$ with an integrated flux above 1 TeV of (1.87 $\pm$ 0.05$_{\mathrm{stat}}$ $\pm$ 0.40$_{\mathrm{syst}}$) $\times$ 10$^{-12} $cm$^{-2}$ s$^{-1}$. The effect of introducing a high cut-off energy in the power law spectrum does not change the integrated flux by more than 10$\%$, less than systematic errors. Integrated fluxes found either with a power law with an exponential cut-off shape or a smoothed broken power law one are consistent with the values given in the previously published H.E.S.S. analysis (Aharonian et al. \cite{PRL}). The spectral parameters for the two curved spectral shapes, for year-wise data, are given in the next two subsections.
\subsubsection{Power law with an exponential cut-off}
\begin{figure}[htbp]
\begin{center}
\resizebox{\hsize}{!}{\includegraphics [width=0.7\textwidth]{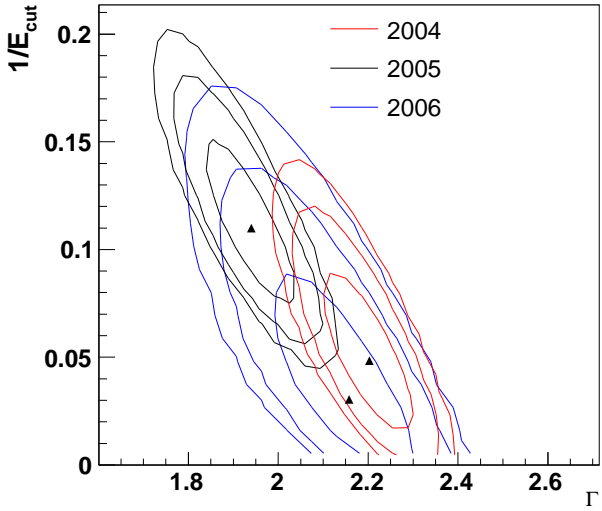}}
\end{center}
\vspace{-0.1cm}\caption{Contours of constant likelihood of the power law with an exponential cut-off fit, as a function of the spectral index $\Gamma$ and the inverse cut-off energy 1/E$_{\rm{cut}}$ for the 2004, 2005 and 2006 datasets. The 68$\%$, 95$\%$ and 99.9$\%$ contours are shown. The triangles denote the best fit positions.}\label{fig11}
\end{figure}

Spectral parameters for the different years (differential flux normalisation $\Phi_0$, spectral index $\Gamma$, cut-off energy and integrated flux above 1 TeV, I($\geq$1 TeV)) are summarized in Table \ref{table2}. The cut-off energy corrected for the systematic bias mentioned in section 3.1 is also shown. The energy ranges of the 2004 and 2006 spectra do not allow the determination of a significant cut-off. The 2004 and 2006 datasets contain fewer runs collected at high zenith angles than the 2005 one. High zenith angle observations provide an increased effective area at high energies and contribute to the high energy part of the spectrum. The 2005 spectral index value is smaller than the values for 2004 and 2006, because of the correlation between the reconstructed spectral index $\Gamma$ and the cut-off energy induced by the fit procedure. Fig. \ref{fig11} shows a 2-D plot of the fitted photon index $\Gamma$ against the cut-off energy for each year's dataset. It can be seen that the spectral parameter values are compatible. Moreover, when fixing the cut-off energy to the uncorrected cut-off energy of 14.7 TeV (see Table \ref{table2}), the spectral fits give $\Gamma$ = 2.14 $\pm$ 0.07$_{\mathrm{stat}}$ $\pm$ 0.10$_{\mathrm{syst}}$, $\Gamma$ = 2.04 $\pm$ 0.04$_{\mathrm{stat}}$ $\pm$ 0.10$_{\mathrm{syst}}$ and $\Gamma$ = 2.11 $\pm$ 0.09$_{\mathrm{stat}}$ $\pm$ 0.10$_{\mathrm{syst}}$ for the 2004, 2005 and 2006 datasets, respectively. Spectral indices are thus compatible with each other.
\begin{table*}[]
\renewcommand\footnoterule{} 
\begin{minipage}[t]{180mm}
\caption{Spectral parameters values with their respective statistical errors for a power law shape with an exponential cut-off, with $\beta = 1$. The fits are performed in the [160 GeV - 70 TeV] energy range.}
\label{table2}
\begin{center}
\begin{tabular}{|c|c|c|c|c|c|c|}
\hline
Year & $\Phi_0$\footnote{Flux normalisation} & $\Gamma$\footnote{Reconstructed spectral index} & E$_{\rm{cut}}$\footnote{Reconstructed cut-off energy without corrections for the systematic bias} & E$_{\rm{cut,true}}$\footnote{Reconstructed cut-off energy corrected for the systematic bias} & I($\geq$1 TeV)\footnote{Integrated $\gamma$-ray flux above 1 TeV} & $\chi^{2}$/dof\\
& $\rm (10^{-12} TeV^{-1} cm^{-2} s^{-1})$ & & (TeV) & (TeV) & $\rm (10^{-12} cm^{-2} s^{-1})$ & \\
\hline 
2004 & 2.40 $\pm$ 0.10 & 2.20 $\pm$ 0.07 & 20.70 $\pm$ 11.80 & 22.20 $\pm$ 11.80 & 1.81 $\pm$ 0.14 & 25/26\\
2005 & 2.56 $\pm$ 0.09 & 1.94 $\pm$ 0.07 & 9.09 $\pm$ 2.13 & 9.61 $\pm$ 2.13 & 2.09 $\pm$ 0.16 & 33/25\\
2006 & 2.35 $\pm$ 0.16 & 2.16 $\pm$ 0.11 & 32.90 $\pm$ 39.50 & 35.50 $\pm$ 39.50 & 1.88 $\pm$ 0.22 & 17/23\\
\hline
\hline
All & 2.55 $\pm$ 0.06 & 2.10 $\pm$ 0.04 & 14.70 $\pm$ 3.41 & 15.70 $\pm$ 3.41 & 1.99 $\pm$ 0.09 & 23/26\\
\hline
\end{tabular}
\end{center}
\end{minipage}
\end{table*}
The spectral shape given by Eq. \ref{Eq2} with $\beta$ = 0.5 is motivated by some shock acceleration scenarios (see discussion in section 5.). A fit over the whole three years with $\beta$ = 0.5 has been performed. It gives a reasonable $\chi^{2}$/dof of 30/23, $\Phi_0$ = (2.55 $\pm$ 0.07$_{\mathrm{stat}}$ $\pm$ 0.40$_{\mathrm{syst}}$) $\times$ 10$^{-12}$ TeV$^{-1}$ cm$^{-2}$ s$^{-1}$, $\Gamma$ = 1.91 $\pm$ 0.08$_{\mathrm{stat}}$ $\pm$ 0.09$_{\mathrm{syst}}$, a low value for the cut-off energy (corrected from the systematic bias) of  E$_{\rm{cut}}$ = (4.0 $\pm$ 1.9$_{\mathrm{stat}}$ $\pm$ 0.7$_{\mathrm{syst}}$) TeV and an integrated flux above 1 TeV of (1.98 $\pm$ 0.04$_{\mathrm{stat}}$ $\pm$ 0.40$_{\mathrm{syst}}$) $\times$ 10$^{-12} $cm$^{-2}$ s$^{-1}$.
\subsubsection{Smoothed broken power law}
Spectral parameters (differential flux normalisation $\Phi_0$, spectral indices $\Gamma_1$ and $\Gamma_2$, break energy and integrated flux above 1 TeV, I($\geq$1 TeV)) are summarized in Table \ref{table3}. The break energy corrected for the systematic bias mentioned in section 3.1 is also given. Parameters show a large dispersion (at most at the 3$\sigma$ level) but remain compatible with each other. The larger value of $\Gamma_2$ for the 2005 data may reflect a possible steepening of the spectrum in the very high energy part.
\begin{table*}[]
\renewcommand\footnoterule{} 
\begin{minipage}[t]{180mm}
\caption{Spectral parameters values with their respective statistical errors for a smoothed broken power law shape. The fits are performed in the [160 GeV - 70 TeV] energy range.}
\label{table3}
\begin{center}
\begin{tabular}{|c|c|c|c|c|c|c|c|l}
\hline
Year & $\Phi_0$\footnote{Flux normalisation} & $\Gamma_1$\footnote{Reconstructed spectral index} & $\Gamma_2^{b}$ & E$_{\rm{break}}$\footnote{Reconstructed break energy without corrections for the systematic bias} & E$_{\rm{break,true}}$\footnote{Reconstructed break energy corrected for the systematic bias} & I($\geq$1 TeV)\footnote{Integrated $\gamma$-ray flux above 1 TeV} & $\chi^{2}$/dof \\
& $\rm (10^{-12} TeV^{-1} cm^{-2} s^{-1})$ & & & (TeV) & (TeV) & $\rm (10^{-12} cm^{-2} s^{-1})$ & \\
\hline 
2004 & 2.39 $\pm$ 0.12 & 2.18 $\pm$ 0.13 & 2.51 $\pm$ 0.23 & 2.40 $\pm$ 0.28 & 2.33 $\pm$ 0.28 & 1.79 $\pm$ 0.58 & 25/19\\
2005 & 2.64 $\pm$ 0.10 & 1.73 $\pm$ 0.10 & 3.07 $\pm$ 0.26 & 3.35 $\pm$ 0.37 & 3.37 $\pm$ 0.37 & 2.11 $\pm$ 0.60 & 28/18\\
2006 & 2.28 $\pm$ 0.18 & 2.27 $\pm$ 0.24 & 2.19 $\pm$ 0.28 & 2.04 $\pm$ 0.38 & 1.94 $\pm$ 0.38 & 1.85 $\pm$ 0.87 & 14/18\\
\hline
\hline
All & 2.57 $\pm$ 0.07 & 2.02 $\pm$ 0.08 & 2.63 $\pm$ 0.14 & 2.61 $\pm$ 0.19 & 2.57 $\pm$ 0.19 & 1.98 $\pm$ 0.38 & 21/19\\
\hline
\end{tabular}
\end{center}
\end{minipage}
\end{table*}
\subsubsection{Spectral variability}
A refined study of the variations of the spectral index with time over the whole three years has been carried out using time intervals of roughly 5 hours, comprising ten consecutive runs. Each data subset has been fitted independently with a power law shape. The spectral index light curve has 25 points. A fit to a constant gives a $\chi^{2}$ of 29/24 d.o.f. confirming that the spectral index did not change significantly over the three years.
\subsection{Correction for absorption of very high energy $\gamma$-rays}
A recent calculation of the Galactic interstellar radiation field (Porter and Strong \cite{Porter}) has shown that the infra-red radiation field near the GC is considerably enhanced compared to what was previously thought. Thus, some attenuation of very high energy $\gamma$-rays might occur at TeV energies. The attenuation coefficient accounting for very high energy $\gamma$-ray absorption by e$^{+}$e$^{-}$ pair production on the interstellar radiation field (ISRF) toward the GC was derived in Zhang et al. (\cite{Zhang}) and can be used to correct the measured spectrum:
\begin{equation}
F(E)=F_0(E)\times\exp(-\tau(E))
\end{equation}
where F(E) represents the observed spectrum after attenuation, F$_0$(E) the intrinsic spectrum and $\tau$(E) the optical depth of the $\gamma$-rays toward the GC as a function of the energy. The optical depth $\tau$(E) depends on the number density of the ISRF photons and on the pair production cross section.\\
Typically, 1$\%$ and 10$\%$ of $\gamma$-rays are absorbed at 10 TeV and 50 TeV, respectively, with an increase of the attenuation factor at higher energies. As expected, the spectral parameters do not change significantly after applying the correction. Thus, the curvature of the spectrum does not seem to be caused by $\gamma$-ray absorption and the cut-off (Eq. \ref{Eq2}) or the spectral break (Eq. \ref{Eq3}) is intrinsic to the source spectrum.
\section{Search for time variability}
As mentioned in the introduction, flaring of Sgr A* and QPOs were detected in various bands such as X-rays and IR. In this section, the ``run by run'' and ``night by night'' light curves are used to search for any variablility in the GC source activity.\\
The run by run light curves are obtained by dividing the data into 28 minutes intervals and by computing the integrated fluxes above 1 TeV for each of these time intervals. The integrated flux is computed from the results of a power law fit with exponential cut-off whose normalization was varied in the likelihood minimization while the spectral index and cut-off energy were fixed to the values obtained for the 2004-2006 dataset. The flux normalisation is adjusted for each 28 minutes time slice. The integrated fluxes above a fixed energy of 1 TeV are then calculated. The night by night light curves are derived in the same way as for the 28 minutes light curves except that the time intervals comprise runs that were recorded in the same night.
\subsection{Light curves and flare sensitivity}
\begin{figure*}[]
\begin{center}
\vspace{0.2cm}
\noindent
\includegraphics [height=0.3\textwidth]{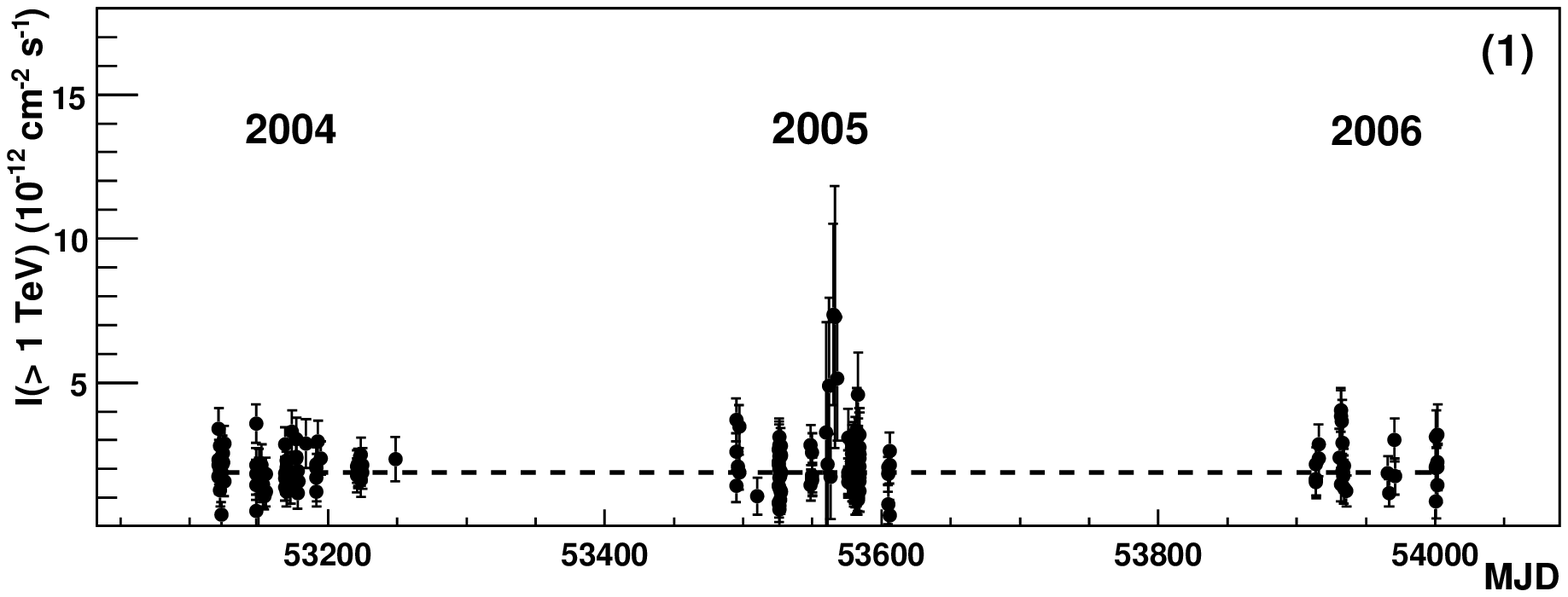}
\includegraphics [height=0.3\textwidth]{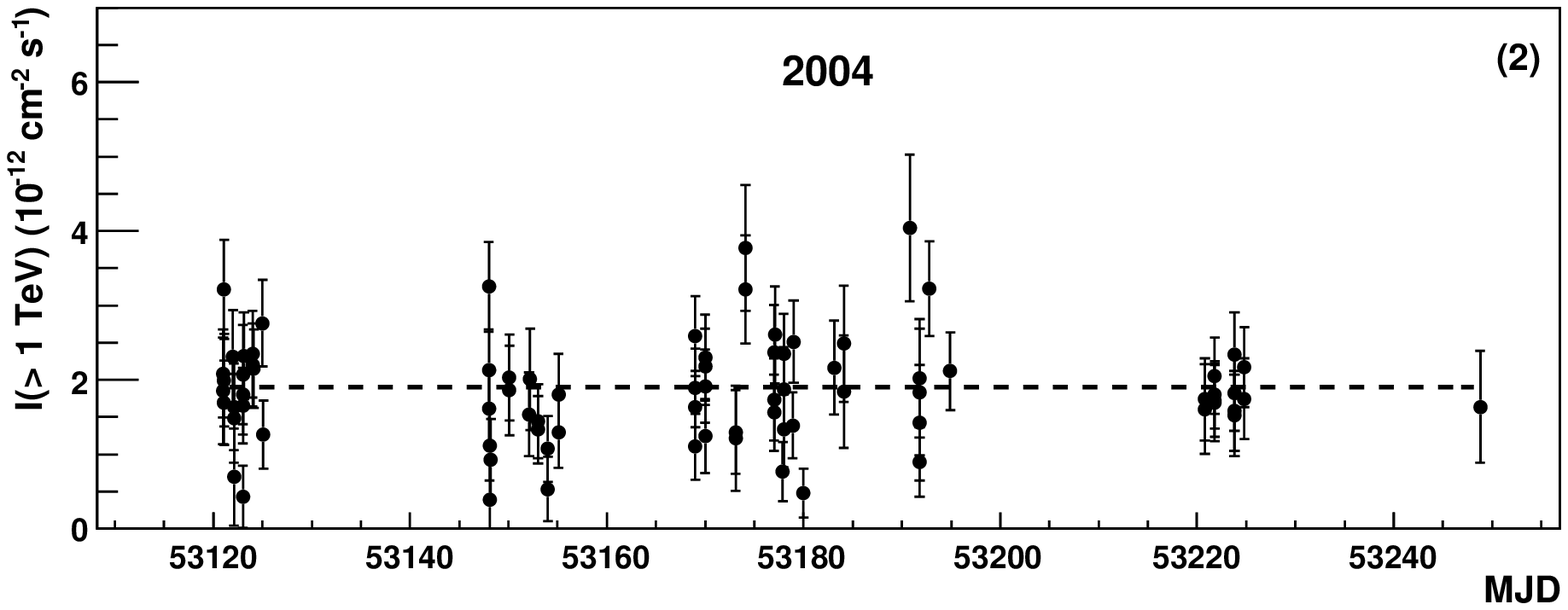}
\includegraphics [height=0.3\textwidth]{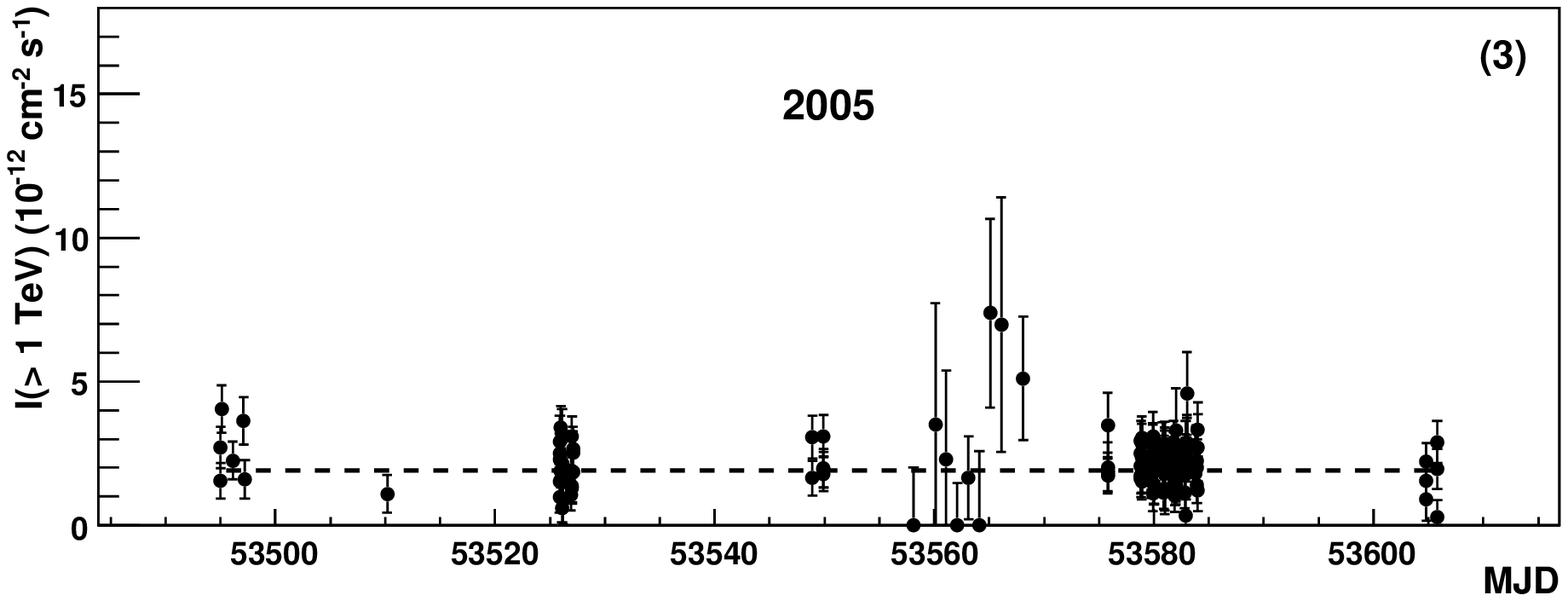}
\includegraphics [height=0.3\textwidth]{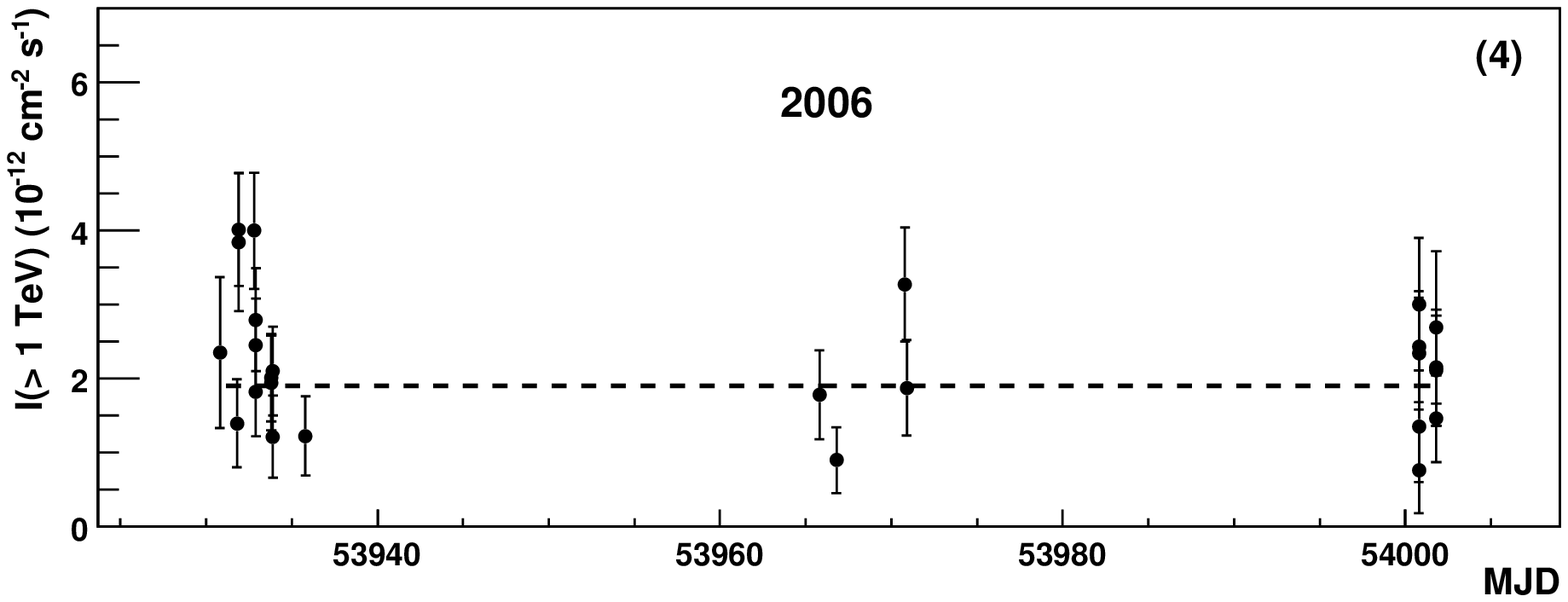}
\end{center}
\vspace{-0.3cm}\caption{Run by run light curves of HESS J1745$-$290. From top to bottom: (1) Data covering the whole 2004-2006 time period. (2), (3) and (4) represent data covering the 2004, 2005 and 2006 time periods, respectively. The dotted lines show the fit to a constant integrated flux of the 2004-2006 lightcurve.}\label{fig1}
\end{figure*}

\begin{figure}[!t]
\begin{center}
\includegraphics [width=0.5\textwidth]{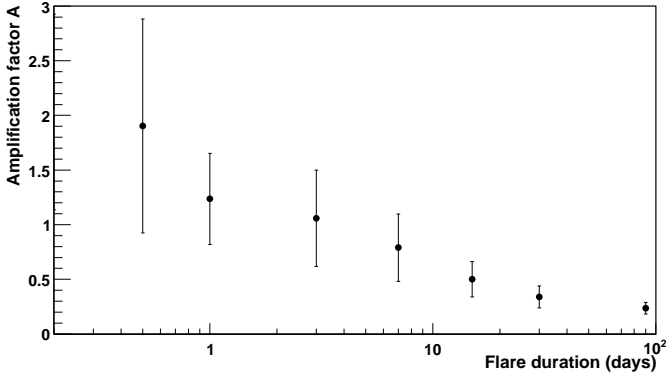}
\end{center}
\vspace{-0.1cm}\caption{Maximum amplification factor $A$ for a 3$\sigma$ flare detection as a function of the flare duration. $A$ is obtained by a Landau fit (see text). The error bars are the corresponding 1$\sigma$ variations, depending on the time of the assumed flare. The maximum amplification factor decreases with increasing flare duration.}\label{fig2}
\end{figure}

The run by run integrated flux light curves of the GC covering the 2004, 2005 and 2006 observation seasons are displayed in Fig. \ref{fig1}. The fit of a constant to the data taken over the whole three years gives a $\chi^{2}$ of 233/216 d.o.f. and does not reveal any variability on time scales longer than 28 minutes.\\
Because of the large error bars of the light curve points implied by the low statistics, the H.E.S.S. signal is only sensitive to relatively large amplitude flares. The flare sensitivity was estimated by adding an artificial Gaussian with variable duration $\sigma_t$, a time of maximum amplification t$_0$ and an amplification at maximum A to the H.E.S.S. light curve (LC):
\begin{equation}
\mbox{LC}_{mod}\mbox{(t)} = \mbox{LC(t)}\times\biggl(1+A\times e^{\frac{(t-t_0)^{2}}{2\sigma_t^{2}}}\biggr).
\end{equation}
When varying the time of maximum amplification t$_0$ along the LC, a distribution of the $A$ values compatible with a flare detection at a given significance is obtained. A fit to this distribution by a Landau function gives the most probable value of $A$ and its corresponding 1$\sigma$ variations. Fig. \ref{fig2} shows the maximum amplification factor $A$ compatible with no flare detection at the 3$\sigma$ confidence level as a function of the flare duration. Long flares involve a larger number of LC points in the constant fit and then increase the $\chi^{2}$ over the number of d.o.f. much more easily, so that $A$ decreases with increasing flare duration. Fig. \ref{fig2} shows that an increase of the flux by at least a factor of two is necessary to detect flares of hour time scales. An increase of the $\gamma$-ray flux of a factor of 2 or greater was excluded at a confidence level of 99$\%$ during a Chandra flare night by the H.E.S.S. collaboration (Aharonian et al. \cite{Hinton}), which is fully consistent with this result.
\subsection{Search for QPOs}
\begin{figure}[!ht]
\begin{center}
\includegraphics [width=0.5\textwidth]{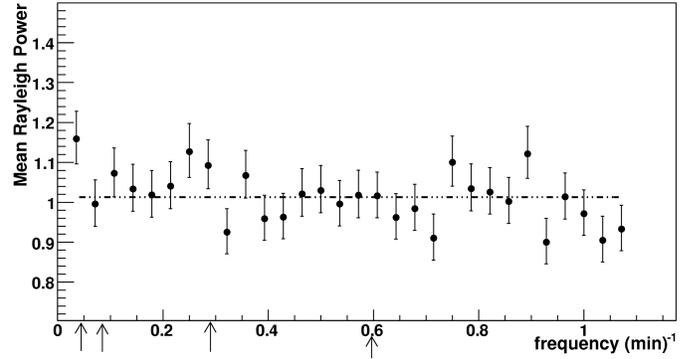}
\end{center}
\vspace{-0.2cm}\caption{Rayleigh power plotted as a function of the frequency. Rayleigh power is normalized such that a pure noise spectrum results in unit power. The dotted line shows the fit to a constant of the Rayleigh spectrum. The arrows denote the 100 s, 219 s, 700 s and 1150 s periods observed in X-rays.}\label{fig3}
\end{figure}

\begin{figure}[!t]
\begin{center}
\noindent
\vspace{-0.3cm}
\mbox{\hspace{-0.1cm}}\includegraphics [width=0.48\textwidth]{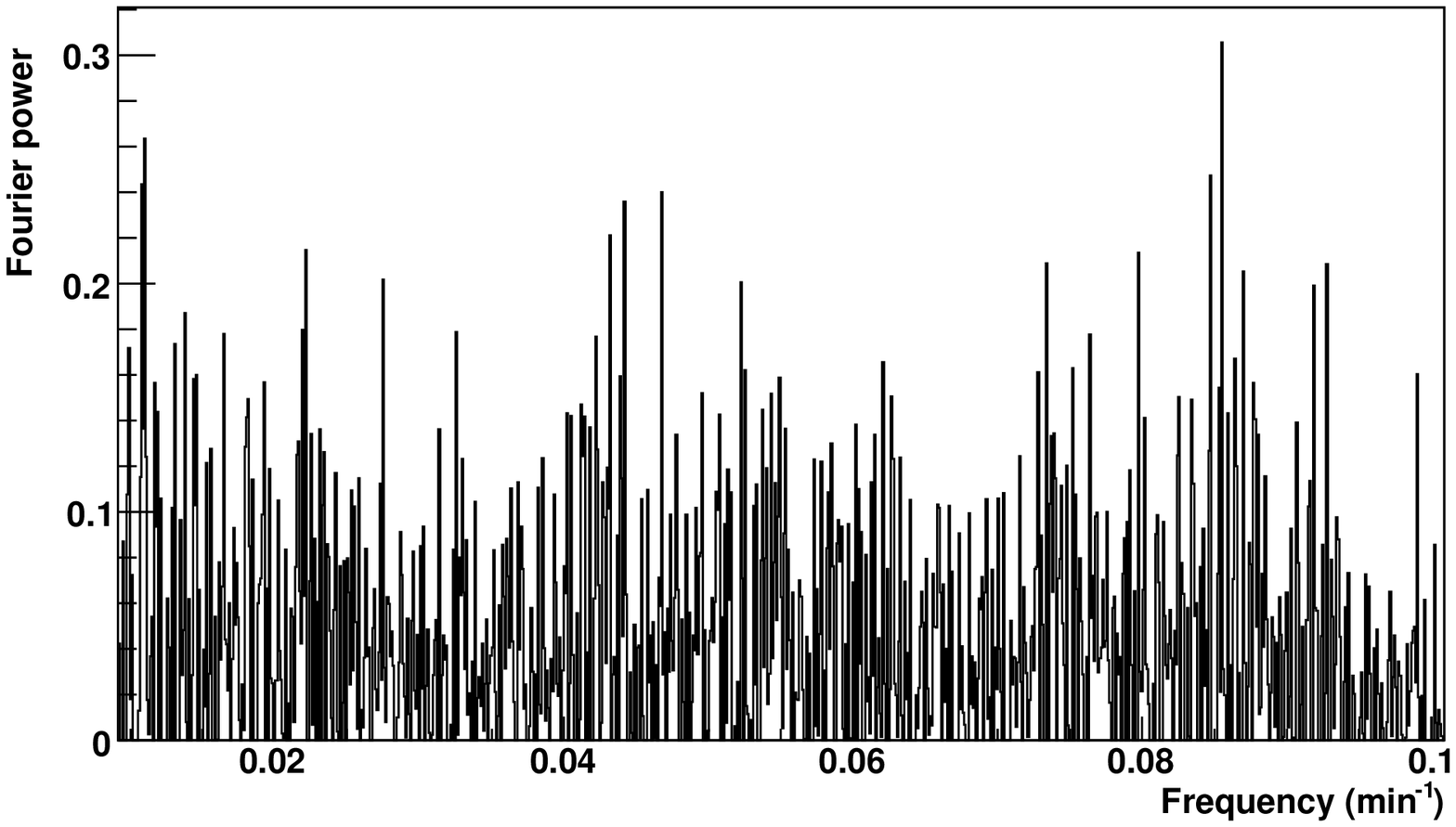}
\includegraphics [width=0.5\textwidth]{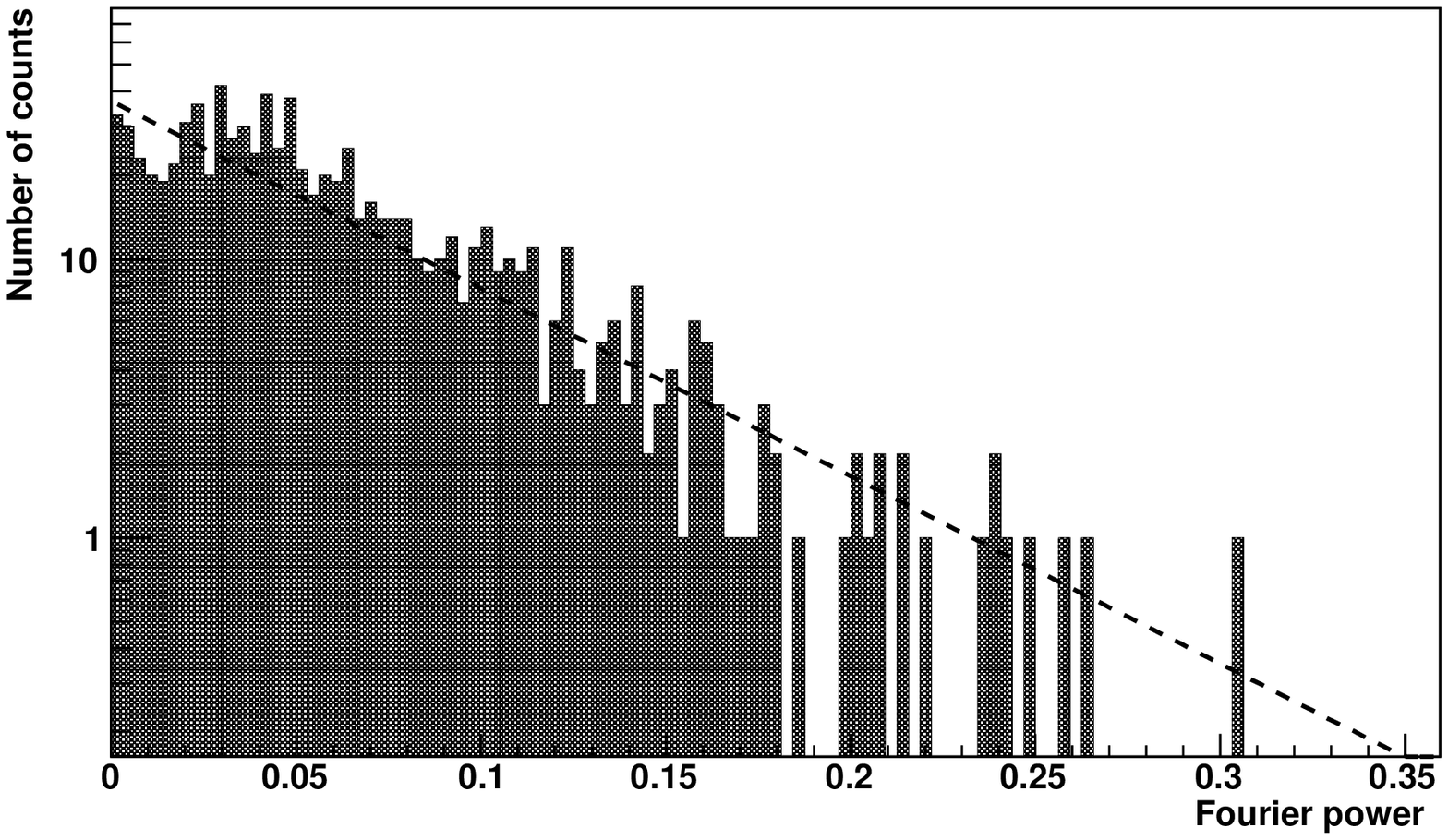}
\end{center}
\vspace{-0.1cm}\caption{Top panel: [10$^{-2}$ min$^{-1}$ - 0.1 min$^{-1}$] Lomb-Scargle periodogram of the H.E.S.S. Sgr A* light curve averaged over the 2004-2006 nights of observation. Bottom panel: Fourier power distribution derived from the averaged Lomb-Scargle periodogram. No significant peak is visible from the Lomb-Scargle periodogram and the $\chi^{2}$ of the exponential fit to the Fourier power distribution is 72/55 d.o.f.}\label{fig4}
\end{figure}

Four oscillation frequencies ranging from 100 s to 2250 s have been observed in the X-ray light curve of Sgr A* (Aschenbach et al. \cite{Aschenbach}). These frequencies, if confirmed, are likely to correspond to gravitational cyclic modes associated with the accretion disk of Sgr A*. The occurence of these frequencies was searched for in the data. First, the coherence time of the disk precession is assumed to be less than 28 minutes. A Rayleigh test (de Jager et al. \cite{deJager}) is then performed on photon arrival time distributions for continuous observations of 28 minutes. Each 28 minutes observation gives a Rayleigh power spectrum. The Rayleigh power averaged over 2004-2006 data is shown in Fig. \ref{fig3} as a function of the frequency. Error bars are estimated by calculating the variance of the Rayleigh power distribution plotted at the corresponding frequency. The probed frequencies range from 1/28 min$^{-1}$ to the inverse of the average time spacing between two consecutive events of 1.2 min$^{-1}$. No Rayleigh power is significantly higher than expected for noise at any probed frequencies. No significant peaks are seen at the 100 s, 219 s, 700 s and 1150 s periods observed in X-rays.
\\In a second analysis, the coherence time of oscillations is assumed to be of the order of a few hours. The Fourier power distribution using Lomb-Scargle periodograms (Scargle \cite{Scargle}) for each night of the dataset is then constructed. Data are binned into 5 minutes intervals. The Fourier power averaged over the 2004-2006 data is displayed in Fig. \ref{fig4} as a function of the frequency. Frequencies tested range from 10$^{-2}$ min$^{-1}$ to 0.1 min$^{-1}$. No significant oscillation frequencies are detected.
\section{Discussion and conclusions}
\begin{figure}[!htpb]
\begin{center}
\resizebox{\hsize}{!}{\includegraphics [width=0.9\textwidth]{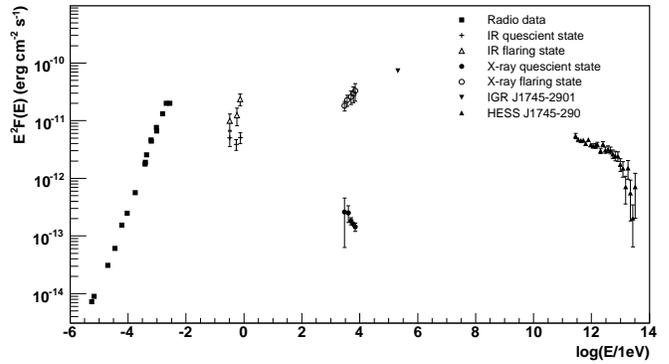}}
\end{center}
\vspace{-0.1cm}\caption{A composite spectral energy distribution of the GC source. The radio data are from Zylka et al. (\cite{Zylka}) and Falcke et al. (\cite{Falcke}). The IR data are from Hornstein et al. (\cite{Hornstein}) and Ghez et al. (\cite{Ghez}). Chandra data points come from Baganoff et al. (\cite{Baganoff} and \cite{BaganoffApJ}), IGR J1745-2901 ones from Belanger et al. (\cite{Belanger}). The HESS J1745-290 spectral points are shown by filled triangles.}\label{fig10}
\end{figure}

A strong signal has been detected from the GC region with the H.E.S.S. instrument. An energy spectrum has been measured with a differential spectrum well described either by a power law with slope $\Gamma$=2.10 $\pm$ 0.04$_{\mathrm{stat}}$ $\pm$ 0.10$_{\mathrm{syst}}$ with an exponential cut-off at 15.7 $\pm$ 3.4$_{\mathrm{stat}}$ $\pm$ 2.5$_{\mathrm{syst}}$ TeV, or a smoothed broken power law with photon indices $\Gamma_1$=2.02 $\pm$ 0.08$_{\mathrm{stat}}$ $\pm$ 0.10$_{\mathrm{syst}}$, $\Gamma_2$=2.63 $\pm$ 0.14$_{\mathrm{stat}}$ $\pm$ 0.10$_{\mathrm{syst}}$ and a break energy at 2.57 $\pm$ 0.19$_{\mathrm{stat}}$ $\pm$ 0.44$_{\mathrm{syst}}$ TeV. An integrated flux above 1 TeV of (1.99 $\pm$ 0.09$_{\mathrm{stat}}$ $\pm$ 0.40$_{\mathrm{syst}}$) $\times$ 10$^{-12} $cm$^{-2}$ s$^{-1}$ is derived using the power law spectrum with an exponential cut-off fit. No indication for variability, flaring or QPOs has been found in the H.E.S.S. data, suggesting a non-variable emission of the GC region in the VHE regime.\\
Different mechanisms have been suggested to explain the broadband spectrum of the GC (Fig. \ref{fig10}). Firstly, the stochastic acceleration of electrons interacting with the turbulent magnetic field in the vicinity of Sgr A*, as discussed by Liu et al. (\cite{Liu}), has been advocated to explain the millimeter and sub-millimeter emission. This model would also reproduce the IR and X-ray flaring (Atoyan and Dermer \cite{Atoyan}). In addition, it assumes that charged particles are accreted onto the black hole, and predicts the escape of protons from the accretion disk and their acceleration (Liu et al. \cite{Liu}). These protons produce $\pi^{0}$ mesons by inelastic collisions with the interstellar medium in the central star cluster of the Galaxy.\\
The cut-off energy found in the $\gamma$-ray spectrum could reflect a cut-off E$_{\mathrm{cut,p}}$ in the primary proton spectrum. In that case, one would expect a cut-off in the $\gamma$-ray spectral shape at E$_{\mathrm{cut}}$ $\simeq$  E$_{\mathrm{cut,p}}$/30. The measured value given in section 3.2.1., with the parameter $\beta$ equal to 0.5, would correspond in this scenario to a low cut-off energy in the primary proton spectrum of roughly 100 TeV. It would correspond to a larger value of $\mathrm{E_{cut,p}}$ of $\mathrm{\sim 400\,TeV}$ if $\beta$ = 1.\\ 
Energy-dependent diffusion models of protons to the outside of the central few parsecs of the Milky Way (Aharonian and Neronov \cite{Neronov2}) are alternative plausible mechanisms to explain the TeV emission observed with the H.E.S.S. instrument. They would lead to a spectral break as in the measured spectrum due to competition between injection and escape of protons outside the vicinity of the GC. A similar mechanism would explain the diffuse emission detected by H.E.S.S. along the galactic plane (Aharonian et al. \cite{DE}).\\
The recent discovery of the G359.95$-$0.04 pulsar wind nebulae (PWN) located at 8'' from Sgr A* also provides interesting models as discussed by Wang et al. (\cite{Wang}) and Hinton and Aharonian (\cite{HintonAharonian}) to explain the steepening of the measured spectrum of HESS J1745$-$290. Inverse Compton emission due to a population of electrons whose energies extend up to 100 TeV might be responsible for at least a fraction of the TeV emission. The PWN model of Wang  et al. and Hinton and Aharonian would imply a constant flux with time since the time scale for global PWN changes is typically much longer than a few years (more like centuries to millennia).
\\The absence of variability in the TeV data suggests that the emission mechanisms and emission regions differ from those invoked for the variable IR and X-ray emission. Moreover, the above-mentioned models can both accomodate a cut-off in the $\gamma$-ray energy spectrum and predict the absence of variability in the TeV emission. They are thus viable scenarios to explain the strong TeV signal detected by H.E.S.S. in the GC region.
\begin{acknowledgements}
The support of the Namibian authorities and of the University of Namibia in facilitating the construction and operation of H.E.S.S. is gratefully acknowledged, as is the support by the German Ministry for Education and Research (BMBF), the Max Planck Society, the French Ministry for Research, the CNRS-IN2P3 and the Astroparticle Interdisciplinary Programme of the
CNRS, the U.K. Particle Physics and Astronomy Research Council (PPARC), the IPNP of the Charles University, the Polish Ministry of Science and Higher Education, the South African Department of Science and Technology and National Research Foundation, and by the University of Namibia. We appreciate the excellent work of the technical support staff in Berlin, Durham, Hamburg, Heidelberg, Palaiseau, Paris, Saclay, and in Namibia in the construction and operation of the equipment.
\end{acknowledgements}

\end{document}